\documentclass[10pt,english,journal]{IEEEtran}
\usepackage{geometry}
\usepackage{iftex}
\ifluatex
  \usepackage{fontspec}
  \usepackage[english, bidi=basic, provide=*]{babel}
  \babelprovide[import, onchar=ids fonts]{english}
  \babelfont{rm}{Noto Sans}
  \usepackage{enumitem}
  \setlist[itemize]{label=-}
\else
  \usepackage[english]{babel}
\fi

\usepackage{amsmath,amssymb,mathtools}
\usepackage{amsthm}
\theoremstyle{definition}
\newtheorem{defn}{Definition}
\usepackage{graphicx}
\usepackage{booktabs}
\usepackage{array}
\usepackage{multirow}
\usepackage{xcolor}
\usepackage{cite}
\usepackage[hidelinks]{hyperref}

\DeclareMathOperator*{\argmax}{arg\,max}

\newcommand{\figph}[2]{%
  \IfFileExists{#1}{\includegraphics[width=\linewidth]{#1}}%
  {\fbox{\parbox[c][2.4cm][c]{0.92\linewidth}{\centering\footnotesize
     \textit{[figure placeholder]}\par\texttt{\detokenize{#1}}\par #2}}}%
}

\newcounter{princtr}
\newcommand{\prinbox}[1]{%
  \refstepcounter{princtr}%
  \begin{center}
  \fbox{\parbox{0.93\linewidth}{\small\textbf{Principle \theprinctr.} #1}}
  \end{center}}

\newbool{suggestions_enabled}
\setbool{suggestions_enabled}{true}

\usepackage[normalem]{ulem}
\newcommand{\carolinasuggestion}[2]{%
  \textcolor{orange}{\sout{#2}}\textcolor{blue}{#1}%
}

\begin{document}

\title{Agents That Model Agents: Five Principles\\ Toward a Theory of Mind for 6G Networks}

\author{
Hatim~Chergui,~\IEEEmembership{Senior~Member,~IEEE}, Carolina Fern\'{a}ndez-Mart\'{i}nez, Mehdi~Bennis,~\IEEEmembership{Fellow,~IEEE}, \\ and Merouane~Debbah,~\IEEEmembership{Fellow,~IEEE}

\IEEEcompsocitemizethanks{\IEEEcompsocthanksitem H. Chergui and C. Fern\'{a}ndez-Mart\'{i}nez are with the i2CAT Foundation, Spain. (e-mails: chergui@ieee.org, name.surname@i2cat.net)}
\IEEEcompsocitemizethanks{\IEEEcompsocthanksitem M. Bennis is with the University of Oulu, Finland (e-mail: mehdi.bennis@oulu.fi).}
\IEEEcompsocitemizethanks{\IEEEcompsocthanksitem M. Debbah is with the Research Institute for Digital Future, Khalifa University, 127788 Abu Dhabi, UAE (e-mail: merouane.debbah@ku.ac.ae).}
\IEEEcompsocitemizethanks{\IEEEcompsocthanksitem This work was supported by grants DGR-2026-2-DIGITALSHIELD, funded by the Research and Universities Department of the Catalonian Government as well as by COALESCE-6G PID2024-163028OB-I00, funded by MICIU/AEI/10.13039/501100011033/FEDER, EU.}
}

\maketitle

\begin{abstract}
Future 6G networks will rely on Large Language Model (LLM) agents to manage
the Radio Access Network (RAN). However, current architectures assume
inter-agent messages convey objective facts. A message is instead a
\emph{trace} of the sender's reasoning: it carries a subjective conclusion,
so a syntactically valid report can propagate an AI hallucination and trigger
a cascading outage invisible to protocol validation. Reading such a trace
requires a Theory of Mind (ToM)---before acting, the receiver must model what the peer believes, and what a peer in that position should have believed.
Modeling these interactions as cognitive channels on a cellular sheaf, we
obtain a unified framework for resilient multi-agent systems, from which five
design principles emerge:
(i) a message is evidence of the sender's hidden reasoning;
(ii) trust is a continuous cognitive Signal-to-Noise Ratio (SNR)---asserted precision over deviation
from the modeled peer belief;
(iii) network-wide consistency and resistance to hallucination contagion are
computable via the sheaf's Laplacian;
(iv) peer-modeling must halt at exactly two levels to conserve compute and
survive mutual information decay; and
(v) credible capacity is bounded by operational goal alignment, not link
bandwidth.
A signaling-storm study on locally deployed 1B-parameter telecom language
models validates it: cognitive SNR isolates a hallucinating peer that three of
its four neighbors agree with, where a divergence gate ranks every wrong peer
above the right one; only depth two ToM recovers the correct action; and the
spectral gap decides whether a topology reaches consistency inside the
near-real-time budget.
\end{abstract}

\begin{IEEEkeywords}
6G, O-RAN, agentic AI society, theory of mind, reasoning, resilience, semantic communication, sheaves, trust.
\end{IEEEkeywords}

\section{Introduction}

\IEEEPARstart{N}{etwork} operators envision 6G infrastructures that read their own telemetry, arbitrate conflicting goals, and autonomously recover from faults without human intervention \cite{tmforum}. The architecture taking shape to fulfill this vision places multiple AI agents inside the Radio Access Network (RAN) Intelligent Controller (RIC). Built on Large Language Models, these agents subscribe to O-RAN E2 measurement reports, exchange semantic messages, and collaboratively decide how the network should behave \cite{agenticran}. When an agent receives a message, it integrates the information into its context window and acts.

This process hides a potentially problematic assumption: that a message is synonymous with a direct measurement, and therefore an intact message implies a sound operational decision. For language-model agents, resilience is no longer merely a transport property. A perfectly delivered message can carry a localized conclusion that inadvertently takes a cell down. What protects the network is the receiver's ability to judge how far to trust the sender's reasoning.

Two critical steps separate the true network state from the final message: what the sender was physically able to observe, and how the sender interpreted that observation. This interpretation step introduces a vulnerability that no current mechanism validates. A message can pass every authentication test yet carry a conclusion that the network reality does not support.

\subsection{A Failure in Five Steps}

For example, a critical failure mode occurs when a massive number of devices attempt to reattach to a neighboring cell during an outage. Telemetry exhibits spiking Radio Resource Control (RRC) attempts and random-access load, while Physical Resource Block (PRB) utilization paradoxically drops (e.g. below 2\%). This PRB collapse stems from signaling congestion, not low demand. Reattaching devices monopolize the control channels with setup messages. This control-plane starvation prevents the scheduler from issuing transmission grants to existing active users. Without these grants, active data bearers stall and are dropped by the congested baseband processor. Unable to allocate resources for user data, the shared data channels empty. User-plane occupancy thus diminishes as control-plane congestion peaks, making low PRB utilization a direct symptom of the storm.

To an anomaly-detection agent trained on user-plane metrics, this inverted correlation is an Out-of-Distribution (OoD) input. Lacking precedent, the agent reverts to learned priors, misclassifies the cell as idle, and cites the accurate PRB measurement to recommend secondary carrier deactivation. The systemic failure materializes downstream. Because this recommendation aligns with an energy-saving agent's utility function, it automatically executes the deactivation, converting transient congestion into a hard outage. No protocol is violated, and the measurement is accurate. The failure originates entirely within the sender's reasoning and propagates because the receiver lacks mechanisms to invalidate the premise.

\subsection{What Is Missing: Theory of Mind}

To contain such failures, receiving agents must reason about \emph{how} messages are produced. In psychology, this is known as Theory of Mind \cite{premack}: the cognitive skill of modeling what someone else believes, including what they might be getting wrong.

Theory of Mind separates two situations that appear identical from the outside. When a peer agent disagrees with a receiver, either the peer is observing valid data that the receiver cannot see (warranting more trust), or the peer has misread shared evidence (warranting less trust). Reacting to the disagreement itself cannot distinguish these realities, and guessing incorrectly drops calls. While machine versions of Theory of Mind exist in multi-agent reinforcement learning \cite{rabinowitz} and LLM-based games \cite{Akata_Schulz_Coda-Forno_Oh_Bethge_Schulz_2025}, there is no unified theory governing them in network control, where decisions occur in milliseconds across multi-vendor agents.

\subsection{A Society of Agents on a Graph}
\label{sec:intro}

We model the control plane as a society of agents positioned on a communication graph $G=(V,E)$. Each agent maintains a local world model, observes a fragment of the network, and repeatedly executes a cycle: it forms a belief, projects that belief to neighbors, weighs their responses, and updates its stance under a hard deadline dictated by the control loop.

This unified construction seamlessly determines several quantities that otherwise require separate treatments: the modeling depth of the peers' beliefs, the required graph connectivity, the rounds needed for consensus, and whether a localized fault cascades.

Consider a single communication link. Let $s$ denote the true network state. No single agent directly observes $s$. Instead, agent $j$ sees local telemetry ${o}_j$, forms an internal belief $b_j$, and emits message $m_j$. The receiver, agent $i$, translates that message into an estimate $Q_i^j$ of $b_j$ before acting $a_i$:
\begin{equation}
\underbrace{s}_{\text{state}} \to
\underbrace{{o}_j}_{\substack{\text{sender's}\\\text{view}}}
\xrightarrow{\ \phi_j\ }
\underbrace{b_j}_{\substack{\text{sender's}\\\text{belief}}} \to
\underbrace{m_j}_{\text{message}} \to
\underbrace{Q_i^j}_{\substack{\text{receiver's}\\\text{estimate}}} \to
\underbrace{a_i}_{\text{action}} .
\label{eq:chain}
\end{equation}
Here, $\phi_j$ represents the sender's latent \emph{type}, and $b_j$ its resulting belief; Principle 1 shows that $b_j$ is the level-0 component of a richer object, the agent's stalk. Because the distortion between state $s$ and message $m_j$ arises from cognitive reasoning rather than physical propagation, \eqref{eq:chain} acts as a \emph{cognitive channel}. Recovering $\phi_j$ from a message history is equivalent to channel estimation, and weighting peers based on inferred reliability is equivalent to signal combining.

\begin{defn}[Harsanyi type]
In game theory involving incomplete information, a player's \emph{type} is a latent variable bundling its private information: payoffs, observation model, and beliefs about other players. Nature draws types from a common prior; players observe only their own. Bayes' rule maps a player's type to a well-defined distribution over others' types \cite{harsanyi}.
\end{defn}

The Harsanyi type terminates an otherwise infinite regress of assumptions (e.g., reasoning about a peer requires a belief about their belief about our state, ad infinitum). Bundling this into one variable creates a finite object over which Bayesian inference operates. For a RAN xApp, the type components are,
\begin{equation}
\phi_j \;=\; \bigl(p_j,\ \beta_j,\ {A}_j,\ \pi_j\bigr) \;\in\; \Phi,
\qquad
\phi_j \sim P_0 .
\label{eq:type}
\end{equation}
These govern how an agent translates telemetry into semantics: $p_j$ is the prior expectation over states; $\beta_j$ collects statistical precisions assigned to priors and evidence; $A_j$ is a mask selecting subscribed E2 counters; and $\pi_j$ is the behavioral policy mapping beliefs to emitted messages. Types are drawn from a codebook $\Phi$ under a prior $P_0$, which represents the operator's catalog of authorized configurations.

The system evolves as a \emph{repeated Bayesian game on a graph}. Each round, agents observe telemetry and emit messages. Receivers update posteriors over neighbors' types using Bayes' rule, estimate actual beliefs, apply trust weights, and fuse data. Because types are never directly observed, the evolving system state is the profile of beliefs about these types.

To govern this network, we utilize Cellular Sheaf Theory. Each agent's local
world model acts as a mathematical \emph{stalk} attached to its node, growing
in dimension with Theory of Mind depth. Links carry projections of that stalk
(\emph{restriction maps}). A sheaf\carolinasuggestion{,}{} rather than a plain weighted graph\carolinasuggestion{,}{} is
required because agents are heterogeneous: subscription masks and reasoning
depths differ, so stalks differ in dimension and no shared state space
exists. The restriction maps absorb this mismatch, letting each link compare
only the vocabulary its two endpoints share. Mismatches between projections
are accumulated over the graph, weighted by trust, and measured by the sheaf
Laplacian \cite{hansen}, whose cohomology further exposes what edge residuals
alone cannot---\emph{hidden} disagreement, where beliefs reconcile on every
link yet admit no consistent global assignment. Resilience is thus quantified
natively as a property of this operator.

\section{Principle 1: A Message Is Evidence of Inference}
\label{sec:problem}

Conventional RAN architectures map state directly to message, ignoring two intermediate links: localized visibility (${o}_j$) and interpretation ($\phi_j$). This reasoning step is precisely where the failures discussed earlier originate.

The latent type introduced in \eqref{eq:type} and the dynamic belief it produces are distinct objects, and this distinction anchors the framework. The type is the agent's internal inference machinery: drawn by Nature, never directly observed by peers, and constant throughout the interaction. The belief is the output generated when that machinery processes current telemetry, and it updates every round. Let $\mathcal{T}_{\phi_j}$ denote the mapping induced by the type. The structured belief state of agent $j$, its \emph{stalk} element, is
\begin{equation}
\begin{aligned}
x_j &= \mathcal{T}_{\phi_j}({o}_j) = \bigl(x_j^{(0)},\, x_j^{(1)},\, x_j^{(2)}\bigr) \in \mathcal{F}(j), \\[2pt]
x_j^{(0)} &= \bigl(\hat\mu_j,\ \Lambda_j\bigr),
\qquad
{\Lambda_j} = \beta_j^{\mathrm{pri}} {I} + \beta_j^{\mathrm{lik}} {A}_j^{\!\top} {A}_j, \\[2pt]
\hat\mu_j &= \Lambda_j^{-1}\bigl(\beta_j^{\mathrm{pri}}\mu_j + \beta_j^{\mathrm{lik}} {A}_j^{\!\top} {o}_j\bigr) .
\end{aligned}
\label{eq:stalk}
\end{equation}

Every symbol traces back to a component of the type $\phi_j$: $\mu_j$ is the mean of the prior $p_j$, while $\beta_j^{\mathrm{pri}}$ and $\beta_j^{\mathrm{lik}}$ are the calibration parameters weighing expectation against new data. The mask $A_j$ selects the E2 counters the agent subscribes to, so telemetry it never reads cannot move its belief. The remaining component, the policy $\pi_j$, maps this belief to an emitted message, closing the generative chain \eqref{eq:chain}.

Here $\Lambda_j$ is a \emph{precision matrix}: the inverse covariance of the agent's belief, and the natural multivariate form of ``how certain am I''. Two features matter operationally. First, precision adds: the prior contributes $\beta_j^{\mathrm{pri}} I$ uniformly, whereas evidence contributes $\beta_j^{\mathrm{lik}} {A}_j^{\!\top} {A}_j$ only along directions the agent can actually observe, so an unsubscribed counter adds no confidence in its direction. Second, the mean is precision-weighted, which is why an agent whose $\beta_j^{\mathrm{lik}}$ collapses is pulled toward its prior. The scalar $\operatorname{Tr}({\Lambda_j})$ used in Principle 2 is simply this confidence summed over all directions \cite{friston2010}.

The superscripts in $x_j$ denote the recursive layers of Theory of Mind, transforming the stalk from a scalar reading into a cognitive model. Level $0$ is the agent's direct belief about the true network state; level $1$ models its belief regarding what a peer believes; level $2$ models its belief about what that peer believes concerning the agent's own state. The stalk space $\mathcal{F}(j)$ is the direct sum of these levels, so $\dim \mathcal{F}(j)$ grows with the depth of recursion---bounded to 2 by Principle 4.

Because vastly different agent configurations can produce identical text outputs, evaluating a message by parsing its surface syntax ($m_j$) is insufficient. Single-message inversion is ill-posed; identification demands a \emph{sequence} of exchanges. The receiver must run backward diagnostic reasoning, estimating the sender's type $\phi$ to infer the network state $s$, maintaining a running posterior over the interaction history $1{:}t$. This derived probability is the peer-induced belief \eqref{eq:qij},

\begin{equation}
Q_i^j(s) \;=\; \sum_{\phi \in \Phi}
\underbrace{P\bigl(b_j\!=\!s \mid m_j, \phi\bigr)}_{\text{invert }\phi\text{'s policy}}
\;\underbrace{P_i\bigl(\phi \mid m_j^{1:t}, o_i^{1:t}\bigr)}_{\text{type posterior from history}} .
\label{eq:qij}
\end{equation}
This Bayes recursion reweights candidate types based on how well their policies explain newly received messages, conditioned on the receiver's local telemetry. Aggregating these probabilities yields a posterior estimate $Q_i^j(s)$ of what agent $j$ genuinely believes.

\prinbox{A peer message $m$ is evidence about the sender's latent type $\phi_j$, and only indirectly about the network state $s$. Evaluating it is a backward inference problem over that hidden type utilizing \eqref{eq:chain}, mathematically well-posed only over a history of messages.}

This framework highlights the inadequacy of standard anomaly detection, which merely evaluates statistical surprise. A surprising but accurate report from an agent with superior visibility should elevate trust; conversely, a highly predictable report stemming from flawed internal reasoning must diminish it. Only the recursive formulation in \eqref{eq:qij} distinguishes between the two scenarios.

\section{Principle 2: Trust Is Cognitive SNR, Not a Gate}
\label{sec:trust}

Receiving a peer's belief estimate ($Q_i^j$ from Principle 1), a node must decide how much to adjust its own belief. Contemporary network designs often treat trust as a binary gate, discarding messages whose divergence exceeds a static threshold. This ignores the asymmetric costs of errors---misclassifying a \texttt{STORM} state as \texttt{IDLE} causes outages, while mistaking \texttt{NORMAL} for \texttt{STORM} merely wastes power. Furthermore, discarding a message extracts zero information, which is mathematically optimal only if the peer holds absolutely no valid data \cite{blackwell}.

Instead, multi-agent networks should process incoming messages as parallel \emph{cognitive channels}. Just as wireless receivers use maximal-ratio combining to weight diversity branches by their Signal-to-Noise Ratio (SNR), a cognitive network maximizes fidelity by weighting each input by its \emph{Cognitive SNR}. Gating is merely the suboptimal extreme of weighting only a single branch.

Formally, let $j$ be the sender, $i$ the receiver, and $e=(i,j)$ the link. Agent $j$ does not transmit its full internal state, but a projection---the \emph{restriction map} $\mathcal{F}_{j\to e}(x_j)$---translating private reasoning into the link's shared vocabulary.

\begin{defn}[Cognitive SNR]
On a link from sender $j$ to receiver $i$, the \emph{signal} is the precision the sender asserts in its transmitted belief. The \emph{noise} is the divergence between that transmitted belief and what receiver $i$ \emph{expected} a sender $j$ to report. The Cognitive SNR reflects confidence discounted by how far it departs from evidence.\footnote{``Cognitive'' here qualifies the reasoning channel of \eqref{eq:chain}, not cognitive radio: $\gamma_{ij}$ is measured over beliefs rather than waveforms, carries both a sender and a receiver index, and is independent of the physical SNR of the underlying link.}
\end{defn}

Let $\Lambda_j$ be the precision matrix of the projected belief, and let $\widehat{b}^{\,(2)}_{\,j|i}$ represent the receiver's expectation of the sender's belief. We express the signal, noise, and resulting weights as,
\begin{equation}
\begin{aligned}
S_j &= \operatorname{Tr}\bigl({\Lambda_j}\bigr),
&
N_{ij} &= D_{\mathrm{KL}}\Bigl( \mathcal{F}_{j\to e}(x_j) \,\big\|\, \widehat{b}^{\,(2)}_{\,j|i} \Bigr), \\[2pt]
\gamma_{ij} &= \frac{S_j}{N_{ij}+\epsilon},
&
w_{ij} &= 1 - e^{-\eta\,\gamma_{ij}} \in [0,1).
\end{aligned}
\label{eq:weight}
\end{equation}
In $\widehat{b}^{\,(2)}_{\,j|i}$, the hat denotes an estimate, $j|i$ means ``$j$ as predicted by $i$,'' and $(2)$ indicates a second-order model ($i$ reasoning about $j$'s reasoning). The signal $S_j$ carries a single index because asserted confidence is intrinsic to the sender. Conversely, noise $N_{ij}$ is link-specific; identical messages are judged differently by receivers with different network models.

The receiver builds $\widehat{b}^{\,(2)}_{\,j|i}$ based on what agent $j$ could \emph{see}, not what it \emph{concluded}. Applying $j$'s public data subscription mask $A_j$ from \eqref{eq:type} (telemetry access metadata), the receiver evaluates that evidence using a neutral prior. This requires modeling the peer's situation independent of its internal reasoning---a true second-order model, and the reason Principle 4 places the floor at two levels. Setting $j=i$ yields the self-weight $w_{ii}$.

Noise (Kullback-Leibler divergence) is evaluated under the operator's cost geometry (Table \ref{tab:cost}), heavily penalizing operationally distant confusions. Parameter $\epsilon$ prevents division by zero, while $\eta$ scales $\gamma_{ij}$'s dynamic range to keep well-separated SNRs distinguishable.

\begin{table}[t]
\centering
\caption{Operator's Cost Geometry: True network state (rows) vs. agent's assumed state (columns). The worst case is normalized to 1. This geometry defines how divergence $N_{ij}$ is measured.}
\label{tab:cost}
\footnotesize
\begin{tabular}{@{}llccc@{}}
\toprule
& & \multicolumn{3}{c}{\textbf{Assumed by the agent}} \\
\cmidrule(l){3-5}
& & IDLE & NORMAL & STORM \\ \midrule
\multirow{3}{*}{\textbf{Truth}}
& IDLE   & 0    & 0.10 & 0.15 \\
& NORMAL & 0.30 & 0    & 0.12 \\
& STORM  & 1.00 & 0.35 & 0    \\ \bottomrule
\end{tabular}
\end{table}

Using easily computable precision as the numerator, this ratio natively acts as a hallucination detector by penalizing divergence rather than raw confidence. An agent facing OoD inputs typically fails in two ways: it reverts to a rigid prior overriding actual evidence (asserting high precision on an unsupported conclusion), or its evidence precision $\beta_j^{\mathrm{lik}}$ collapses while still committing to a confident categorical answer. In either case, the receiver expects a peer observing unfamiliar telemetry to report differently, causing $N_{ij}$ to spike, so the discount does not depend on which mode occurred. The first mode mirrors computational psychiatry, where hallucinations occur when dominant internal priors override sensory evidence \cite{adams}.

Uninformative peers assert no precision, driving $S_j \to 0$ and reproducing a ``reject'' gate without hard thresholds. To prevent rogue agents from inflating confidence to dominate decisions, long-term reputation is graded by a \emph{strictly proper scoring rule} \cite{gneiting}, preserving influence only for those honest about uncertainty.

\prinbox{Trust is a fluid, continuous variable measured as the peer's Cognitive SNR \eqref{eq:weight}: the precision asserted divided by how far that assertion departs from a second-order Theory of Mind expectation. Confidence earns trust only insofar as the evidence warrants it---unjustified certainty and collapsed precision alike are heavily discounted.}

\section{Principle 3: Agent-Society Topology Governs Consistency and Contagion}
\label{sec:society}

Principles 1 and 2 concern a single receiver reading a single peer, and yield one number per link: the trust weight $w_{ij}$. A live RAN control plane consists of many such channels, and those weights turn out to be exactly what the network-scale analysis needs. Do localized beliefs compose into a globally consistent state, and how quickly? Cellular sheaf theory answers both.

For any edge $e=(i,j)$, the disagreement between the two agents' projections is the coboundary,
\begin{equation}
\delta(x)_e \;=\; \mathcal{F}_{i\to e}(x_i) \;-\; \mathcal{F}_{j\to e}(x_j),
\label{eq:sheaf}
\end{equation}
where $x=(x_1,\dots,x_{|V|})$ collects the stalk elements \eqref{eq:stalk} of all agents, and $\mathcal{F}_{i\to e}$ is the same restriction map used in \eqref{eq:weight}. Assembling $\delta$ over every edge yields a linear map whose kernel and cokernel carry direct operational significance:
\begin{equation}
H^0 = \ker \left(\delta\right),
\quad
H^1 = \operatorname{coker}\left(\delta\right).
\label{eq:cohomology}
\end{equation}

$H^0$, the space of global sections, contains the belief assignments on which every pair of neighbors agrees; it defines the achievable consensus. $H^1$, the obstruction space, contains underlying disagreements no local adjustment can erase. For scalar stalks, $\dim H^1$ reduces directly to the cycle rank of the graph ($\dim H^1 = |E| - |V| + 1 $). Each independent loop contributes one degree of freedom where pairwise agreement fails to guarantee global consistency. A tree topology has zero cycle rank, meaning pairwise verification is globally sufficient---the cognitive equivalent of Kirchhoff's voltage law.

Because $\delta$ is defined edge-by-edge, this operation applies seamlessly to a single link, a cluster, or the entire RAN graph. This perfectly matches O-RAN's hierarchical architecture, providing fast local tests in the near-real-time loop and global tests in the non-real-time loop.

The network's convergence speed follows from the sheaf Laplacian $L_{\mathcal{F}} = \delta^{\!\top}\delta$. Iterative belief exchange reduces the residual $\lVert\delta(x)\rVert^2$ at a rate strictly governed by $\lambda_2$ (its smallest nonzero eigenvalue). Time to consistency scales as $\tau_{\text{cons}} \propto 1/\lambda_2$. Because edge weights are strictly the cognitive SNRs from \eqref{eq:weight}, trust and consistency speed are the same physical quantity.

Two spectral failure modes emerge. If trust collapses across enough links, the graph approaches mathematical disconnection ($\lambda_2 \to 0$), and $\tau_{\text{cons}}$ diverges. The society never achieves consistency; localized hallucinations are not corrected but spread systemically \cite{pastor}. Conversely, sparse topologies with healthy trust inherently have small spectral gaps, reaching consistency too slowly. The design mandate is singular: keep $\lambda_2$ above the critical threshold required by the latency budget.

\prinbox{The coboundary \eqref{eq:sheaf} makes local-to-global consistency computable at any network scale. Its cohomology \eqref{eq:cohomology} cleanly separates achievable consensus ($H^0$) from irreconcilable disagreement ($H^1$). The spectral gap of the associated Laplacian dictates the time to consistency and the network's resistance to hallucination contagion, using cognitive SNRs as edge weights.}

In dense meshes, it is critical that at least one agent per loop is anchored directly to verifiable E2 hardware counters to close the cognitive cycle against physical ground truth.

\section{Principle 4: Recursion Stops at Two Levels}
\label{sec:depth}

Principle 3 left one quantity unpinned: the dimension of the stalk, which enters the Laplacian and therefore the spectral gap. That dimension is set by how many levels of Theory of Mind each agent maintains, so the design question becomes how deep to recurse. In multi-vendor RICs, proprietary boundaries prevent xApps from inspecting each other's algorithms. Stable cooperation rests on modeling both the peer's network beliefs and the peer's beliefs regarding the receiver's beliefs, mirroring sociological double contingency \cite{luhmann}. In the stalk of \eqref{eq:stalk} this is the superscript: depth $d$ means retaining the levels $x^{(0)},\dots,x^{(d)}$.

This recursion cannot proceed indefinitely. The optimal depth $d^{\star}$ balances the strategic advantage $V(d)$ against the computational cost $C(d)$ dictated by the near-real-time latency budget $\ell$, i.e.,
\begin{equation}
d^{\star} \;= \; \argmax_{d \in \mathbb{N}}\ \bigl\{\,V(d) - \ell\, C(d)\,\bigr\} .
\label{eq:depth}
\end{equation}
Two realities pin this depth to exactly two levels.

\emph{One level is not enough.} A level-1 receiver reads assertions but cannot estimate distortion. Calibrating cognitive SNR \eqref{eq:weight}, however, requires a baseline expectation of what the peer \emph{ought} to have said, demanding a second-order model. Level 2 is the minimum depth at which trust becomes computable.

\emph{Three levels are too many.} Recursive modeling forms a Markov chain of lossy cognitive channels where the data-processing inequality applies,
\begin{equation}
X \to Y \to Z \to W
\;\Longrightarrow\;
I(X;Y) \ge I(X;Z) \ge I(X;W).
\label{eq:dpi}
\end{equation}
This dictates a monotone decay in mutual information. The distinction lies in physical anchors. Level-2 models terminate in data the receiver can verify via local E2 telemetry, yielding falsifiable predictions. The mathematical decay is actively opposed by physical measurement.

Level 3 admits no such physical anchor. It concerns what the peer believes about the receiver's beliefs---a purely psychological quantity E2 counters cannot observe. The surviving mutual information crashes into the noise floor, driving cognitive SNR and corresponding edge weights to zero. Through \eqref{eq:cohomology}, the graph mathematically disconnects, irreparably fragmenting the consensus space $\dim H^{0}$. Level 2 is the exact sweet spot: deep enough to calibrate trust, but shallow enough to remain tethered to reality \cite{camerer}.

\prinbox{Model the peer, and the peer's model of the receiver---go no deeper. Level 2 is the minimum depth at which cognitive SNR is mathematically computable, and the absolute maximum depth whose predictions remain anchored by physical telemetry. A third level lacks observable anchors, drives network edge weights to zero, and irreparably fragments the consensus space.}

\begin{figure}[t]
\centering
\figph{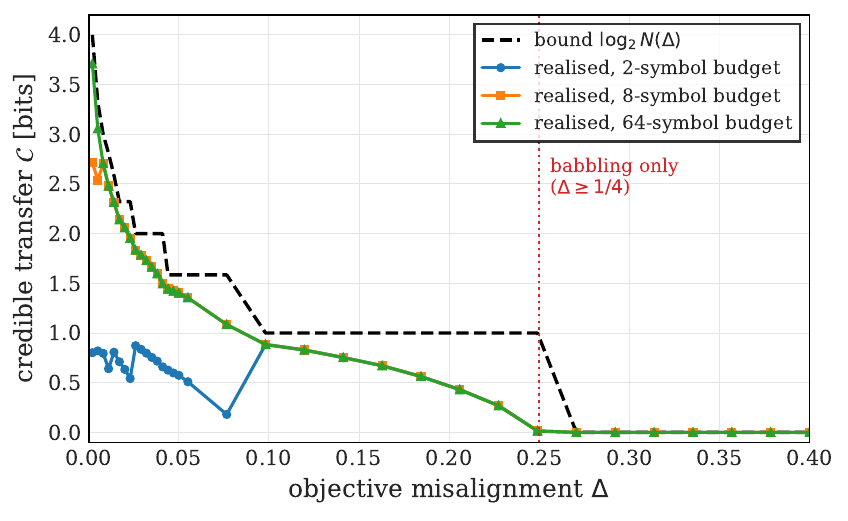}{Information that gets through between two agents versus how far apart their goals are, for three control-channel budgets.}
\caption{How much two agents can actually tell each other, as their operational goals drift apart. The credible information curve crashes to zero even while the physical transport channel remains wide open.}
\label{fig:capacity}
\end{figure}

\section{Principle 5: Alignment, Not Bandwidth, Bounds Communication}
\label{sec:align}

Principles 1 to 4 bound what agents can \emph{infer} about one another. A final limit is independent of inference entirely and arises from what agents \emph{want}: even a perfectly calibrated receiver facing a perfectly capable sender learns little if their objectives diverge. We therefore establish a bound on the capacity of the cognitive channel itself. When agents optimize competing utilities, their control-plane messages are costless and unverifiable \emph{cheap talk} \cite{crawford}. Indeed, Crawford and Sobel demonstrated that when a sender and a receiver hold divergent objectives, the sender structurally cannot communicate fine-grained, credible information. To formalize this, let $\Delta$ denote a scalar misalignment parameter that quantifies the geometric divergence between the agents' utility functions (i.e., the distance between their preferred optimal actions for a given state).

\begin{figure*}[t]
\centering
\figph{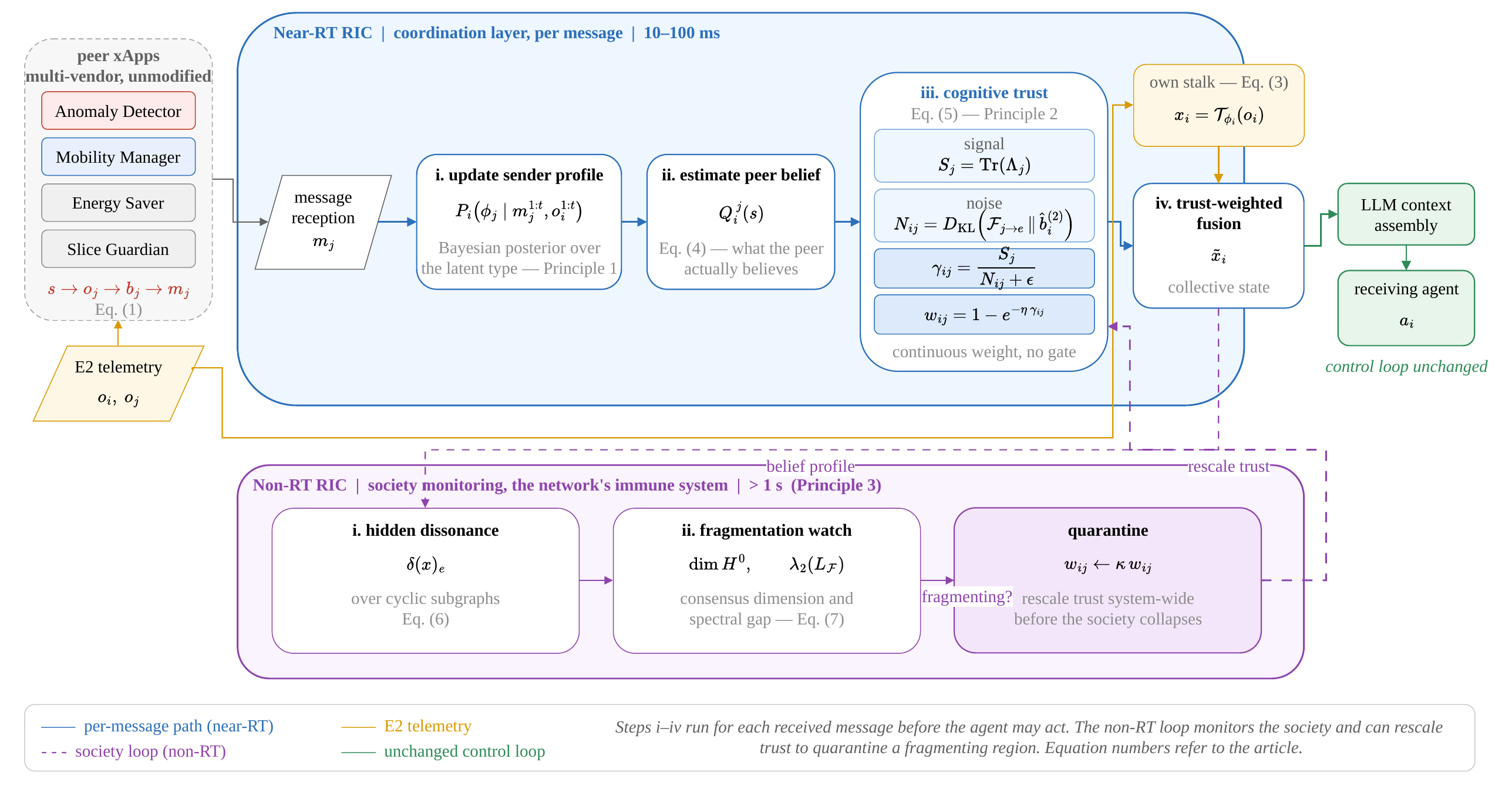}{The peer-modeling layer sits between message reception and context assembly in the near-real-time RIC, with society-wide statistics kept at the non-real-time RIC.}
\caption{The coordination layer inside a functional RIC, on two timescales. Per message, the near-real-time loop runs four steps before the receiving agent may act: update the posterior over the sender's latent type, estimate the peer belief $Q_i^j$, evaluate the cognitive SNR trust weight $w_{ij}$, and fuse the trust-weighted belief with the receiver's own. In parallel, the non-real-time loop monitors the society, tracking hidden dissonance through the coboundary $\delta(x)_e$ and watching $\dim H^0$ for fragmentation, and can rescale trust system-wide to quarantine an affected region. Proprietary agents and the underlying control loop are unmodified.}
\label{fig:arch}
\end{figure*}

Because the sender is incentivized to skew the receiver's actions toward its own utility, it will naturally exaggerate its reports. The receiver, being a rational actor, mathematically anticipates this strategic bias. Consequently, highly precise, granular claims lose all credibility, as the receiver knows the sender has a constant motive to shade exact numbers. To establish any believable communication, the interaction must structurally retreat into coarse categorizations, or ``bands'' of truth. These intervals must be sufficiently wide that the sender cannot profitably lie across boundaries without triggering an outcome that severely penalizes its own utility.

The mathematical consequence is that the resolution of credible information strictly degrades as the misalignment of objectives increases. The maximum credible information transfer, $\mathcal{C}$, in bits, is bounded by,
\begin{equation}
\mathcal{C} \;\le\; \log_2 N(\Delta),
\qquad
N(\Delta) = \left\lceil -\tfrac{1}{2} + \tfrac{1}{2}\sqrt{1+\tfrac{2}{\Delta}}\,\right\rceil ,
\label{eq:capacity}
\end{equation}
where $N(\Delta)$ defines the maximum number of distinct operational bands the sender can credibly convey. For example, if the misalignment $\Delta$ across utility functions is minimized, the sender might reliably distinguish among four granular states (e.g., ``critical,'' ``high,'' ``moderate,'' and ``low''). As the divergence $\Delta$ grows, the overwhelming temptation to exaggerate forces the credible bands to widen, collapsing the resolution to two coarse states (e.g., ``high'' or ``low'').

This formulation establishes a fundamental limit: objective divergence inherently caps the granularity of communication. Once $\Delta$ reaches $1/4$, $N(\Delta)$ reduces to 1 and the usable capacity is exactly zero. Beyond this threshold, the sender's bias is so extreme that it will strategically output the exact same message regardless of the true underlying state. At this point, the messages cease to convey any real information.

\prinbox{The credible information conveyed between two agents \eqref{eq:capacity} is strictly bounded by the alignment of their operational objectives. Because rational receivers anticipate exaggeration, precise communication loses granularity. Beyond a critical threshold of misalignment, this informational capacity reaches zero; agents can exchange syntactically perfect messages indefinitely while conveying no usable data. Therefore, objective alignment---not physical link capacity---is the fundamental bottleneck to trusted agentic coordination.}
\vspace{0.5cm}
\section{Putting It Together}
\label{sec:arch}

To see how these principles operate in practice, Figure \ref{fig:arch} illustrates their integration into a functional RIC. Rather than requiring massive architectural overhauls, these principles are implemented as a lightweight coordination layer explicitly positioned between the raw reception of a message and the LLM context assembly. This strategic placement leaves the underlying control loops fully intact and requires zero modification to the proprietary, multi-vendor agents themselves.

The system operates seamlessly across two distinct timescales: a rapid, per-message evaluation loop and a slower, society-wide monitoring loop. During the fast per-message evaluation, the coordination layer intercepts incoming communications and executes four mathematically grounded steps before the receiving agent is allowed to act: i) it updates the sender's profile (Principle 1) by using the new message to refine a running Bayesian posterior probability over the sender's latent type; ii) it translates this updated profile into a concrete estimate of what the sender actually believes about the network state, represented as $Q_i^j$ in \eqref{eq:qij}; iii) it evaluates cognitive trust (Principle 2) by synthesizing the precision the sender asserts against how far that assertion deviates from expected behavior, outputting the continuous cognitive SNR trust weight $w_{ij}$ via \eqref{eq:weight}; and iv) it smoothly blends this trust-weighted peer belief with the receiver's own internal priors. Only this stabilized, trust-filtered collective state is passed to the agent to drive a network action.

Simultaneously, background processes running in the slower non-real-time RIC act as the network's immune system (Principle 3) by monitoring the overall health of the agent society. This slower loop performs two vital functions: i) it tracks hidden cognitive dissonance via a continuous consistency check that evaluates the coboundary \eqref{eq:sheaf} over cyclic subgraphs; and ii) it prevents systemic failure by actively tracking the consensus dimension $\dim H^{0}$ of \eqref{eq:cohomology}. If localized hallucinations begin to spread and the society shows signs of mathematically fragmenting, the non-real-time RIC can proactively intervene, rescaling trust weights system-wide to quarantine the affected agents before the network collapses.

\section{Case Study}
\label{sec:case}

We simulate a near-real-time RIC governing a single cell, wherein the network telemetry undergoes a severe transition from a quiescent \texttt{idle} state to a massive signaling \texttt{storm}. The multi-agent society operating on this RIC comprises five distinct xApps, each instantiated as a locally deployed 1-billion-parameter telecom Small Language Model (SLM), denoted as \texttt{otellm}. Each agent is prompted strictly with its specific operational role and its subscribed E2 Service Model, Key Performance Measurements (E2SM-KPM) counters. At every timestep, an agent infers the current network regime and outputs a confidence score. This intrinsically yields a level-$0$ belief and an asserted precision, $S_j$, derived directly from the model rather than relying on externally imposed heuristics. Crucially, an agent's capacity to recognize a given network regime is strictly bounded by its prior training exposure. To ensure methodological rigor, our evaluation isolates distinct system dynamics: while large-scale topology and latency sweeps (Fig.~\ref{fig:envelope}) utilize a calibrated statistical emitter to decouple multi-agent coordination costs from SLM decoding variance, all empirical decision-quality metrics (Tables~\ref{tab:results} and~\ref{tab:depth}) are generated entirely by the deployed \texttt{otellm} models.

Within this framework, our primary scenario introduces a deliberate informational asymmetry. Four user-plane agents---the Traffic Balancer (acting as the primary receiver), the Energy Saver, the Slice Guardian, and the Anomaly Detector---lack prior training exposure to signaling storms. As the storm manifests, these four agents misinterpret the sudden drop in PRB utilization as a symptom of low user demand, erroneously classifying the network state as \textsc{idle}. Consequently, the receiver draws an incorrect conclusion from its own local evidence, an error that is subsequently reinforced by three of its peers. Conversely, the Mobility Manager, which monitors Random Access Channel (RACH) and RRC counters and possesses prior training on storm regimes, correctly diagnoses the event. This configuration constructs a highly adversarial consensus environment: to achieve accurate state resolution, a correct minority of one must overcome not only a flawed majority consensus, but also the receiver's own erroneous initial belief.

\subsection{Trust Discriminates Where Disagreement Cannot}

Table~\ref{tab:results} details peer evaluation by the Balancer at storm onset. Because all peers comparably displace its initial belief, naive disagreement metrics fail to identify the correct agent.

For the deployed SLM, the Anomaly Detector's failure manifests as a collapse in likelihood precision rather than loud overconfidence. Encountering OoD storm telemetry, it asserts low precision while still erroneously reporting \texttt{idle}. Our trust mechanism captures this via the noise term---as a calibrated peer would not report \textsc{idle} under these conditions---driving its cognitive SNR down to $-4.3$\,dB.

Conversely, the storm-trained Mobility Manager legitimately asserts high precision ($+7.8$\,dB). This $12$\,dB spread over the mistaken majority yields a thirteenfold advantage in trust weight, exposing a critical flaw in naive consensus: the \emph{correct} peer diverges \emph{most} from the group. Consequently, a standard divergence gate actively prefers erroneous agents Area Under the Curve $\text{(AUC)} = 0.000$. By evaluating cognitive SNR instead of raw divergence, our mechanism perfectly isolates the accurate peer ($\text{AUC} = 1.000$). Applying this internally, the receiver self-assigns a low trust weight ($0.17$), enabling the correct minority to overcome both a flawed majority and its own erroneous belief.

\begin{table}[t]
\centering
\caption{Cognitive SNR at storm onset, averaged over 80 rounds.}
\label{tab:results}
\footnotesize
\begin{tabular}{@{}lcccc@{}}
\toprule
\textbf{Peer} & $S_{j}$ & $N_{ij}$ & $\gamma_{ij}$ \textbf{[dB]} & $w_{ij}$ \\ \midrule
Anomaly Detector & 2.93 & 11.950 & $-4.3$ & 0.036 \\
Mobility Manager & 40.00 & 6.676 & $\mathbf{+7.8}$ & 0.451 \\
Slice Guardian & 3.38 & 2.720 & $+3.6$ & 0.159 \\
Energy Saver & 3.18 & 2.600 & $+4.0$ & 0.172 \\
\bottomrule
\end{tabular}
\end{table}

\subsection{Two Levels, and No More}

Table~\ref{tab:depth} sweeps the recursion depth of reasoning. Depth $0$ ignores peers entirely, inheriting the Balancer's own error. Depth $1$ incorporates peers but lacks a baseline expectation to weigh them against, reducing trust to a blind guess where the wrong majority wins. Depth $2$ successfully recovers the correct action in every trial. Depth $3$, however, collapses to near-guessing accuracy while costing significantly more. This failure occurs because the third level attempts to calculate what a peer believes about the receiver's \emph{own} beliefs---an abstract quantity unobserved by any E2 counter, causing surviving mutual information to decay into the noise floor. Crucially, our timing measurements reveal the computational bottleneck is not the dimensionality of the mathematical belief object, but the nested second-order evaluation. Therefore, two levels of reasoning provide the exact optimal balance of accuracy and computational efficiency.

\begin{table}[t]
\centering
\caption{Recursion depth against decision quality and measured cost. Timings are medians over 5 runs of 60 messages, covering the full nested evaluation path.}
\label{tab:depth}
\footnotesize
\begin{tabular}{@{}ccccc@{}}
\toprule
\textbf{Depth} & $\dim\mathcal{F}(j)$ & \textbf{Cost} & \textbf{Decision} & \textbf{Calibration} \\
$d$ & \textbf{(stalk)} & \textbf{[ms/msg]} & \textbf{accuracy} & \textbf{error} \\
\midrule
0 & 1 & 0.32 & 0.000 & 0.955 \\
1 & 5 & 0.40 & 0.050 & 0.909 \\
2 & 21 & 1.52 & \textbf{1.000} & 0.676 \\
3 & 85 & 2.57 & 0.083 & 0.771 \\
\bottomrule
\end{tabular}
\end{table}

\subsection{Does the Society Converge in Time?}

\begin{figure}[t]
\centering
\figph{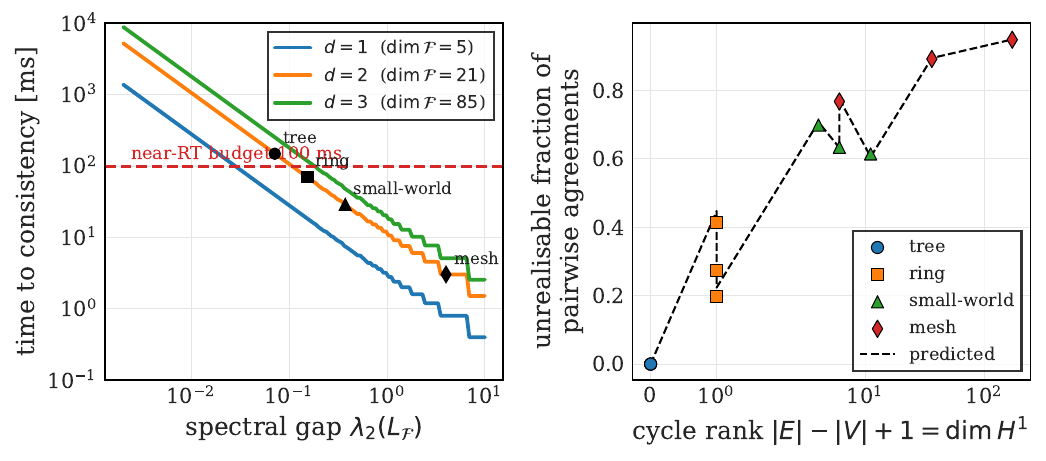}{Left: time to consistency against spectral gap for each recursion depth, with the near-real-time budget marked. Right: unrealizable fraction of pairwise agreements against cycle rank.}
\caption{The design envelope. Left: time to consistency scales with the network's spectral gap while per-message cost is set by the nested second-order evaluations each depth requires, so depth, connectivity and trust trade against one another under a fixed latency budget. Right: hidden disagreement is impossible on trees and grows with cycle rank.}
\label{fig:envelope}
\end{figure}

The ultimate operational question is whether consistency is reachable before the $100$\,ms near-real-time control loop closes. Fig.~\ref{fig:envelope} (left) plots time-to-consistency against the spectral gap. At depth $2$ and moderate trust, topology dictates performance: a tree misses the budget ($149$\,ms), a ring succeeds ($70$\,ms), a small-world graph accelerates to $29$\,ms, and a full mesh converges in just $3$\,ms. Connectivity, trust, and reasoning depth are not independent; they trade against one another through a single measurable scalar.

Fig.~\ref{fig:envelope} (right) confirms that while trees exhibit zero hidden disagreement, denser topologies harbor unrealizable agreements that grow alongside cycle rank. Finally, communication capacity behaves exactly as predicted: once operational misalignment between agents reaches $25\%$, usable capacity collapses to zero. Widening the communication alphabet yields no benefit once goal alignment, rather than bandwidth, becomes the binding constraint (Fig.~\ref{fig:capacity}).

\section{Open Problems}
\label{sec:open}

Transitioning this resilience framework from mathematical models to production-grade O-RAN deployments introduces several practical challenges:
(i) \textbf{Cognitive Channel Estimation:} Agents must quickly learn how their peers reason. Future work must quantify the ``pilot-like'' signaling overhead this continuous estimation imposes, ensuring agents can learn peer behaviors without choking the live control plane;
(ii) \textbf{Eliciting Calibrated Precision:} Our agents failed when their categorical decisions contradicted their collapsing asserted precision. Reliably extracting this precision ($S_j$) from LLMs at line rate---across decoding temperatures and prompt formats---remains an open challenge, as does adapting theoretical communication limits to their bounded rationality;
(iii) \textbf{Real-Time Society Tracking:} To prevent ``systemic hallucinations,'' the network must continuously monitor the agent society's health. This requires ultra-lightweight statistical tools capable of processing execution logs within the strict microsecond-to-millisecond latency budgets of the near-real-time RIC;
(iv) \textbf{Translating Operator Intents:} Our risk-weighted trust mechanism requires precise numerical penalties. Because operators express priorities through qualitative Service Level Agreements (SLAs), the industry needs standardized frameworks to automatically translate these high-level business intents into concrete cost matrices; and
(v) \textbf{Hardware-in-the-Loop Validation:} This coordination layer requires rigorous benchmarking on physical O-RAN testbeds to evaluate true end-to-end inference latency, memory footprint, and E2 signaling overhead under live, chaotic radio traffic patterns.

\section{Conclusion}
\label{sec:conclusion}

An agentic 6G RAN operates as an artificial society where valid data can easily propagate flawed conclusions. Network resilience thus relies on continuous, mathematically calibrated trust rather than simple link integrity. Securing these control planes requires equipping the network with a Theory of Mind to evaluate peer reasoning. By treating inter-agent communication as a cognitive channel distorted by subjective reasoning, five principles govern this society:
(i) read messages as evidence of hidden types, keeping operational uncertainty explicit via distributions;
(ii) scale trust using a continuous cognitive SNR to natively discount hallucinated reports without relying on separate anomaly detectors;
(iii) model the agent society as a mathematical sheaf, using its Laplacian for consistency and its spectral gap to bound consensus speed;
(iv) cap cognitive recursion at two levels, the depth where trust remains computable and survives mutual information decay; and
(v) prioritize objective alignment over raw bandwidth, since misaligned operational goals collapse communication capacity to zero.
Operationally, across a society of locally deployed telecom SLMs, this framework recovered the correct action even when a naive divergence gate preferred every mistaken peer over the right one. Ultimately, it unifies reasoning depth, connection density, and consensus speed into a single measurable spectral quantity.



\bibliographystyle{IEEEtran}
\bibliography{refs}

\end{document}